\documentclass[pra,aps,english,twocolumn,notitlepage,superscriptaddress]{revtex4-2}
\usepackage[T1]{fontenc}  
\usepackage[latin1]{inputenc}
\usepackage{multirow,amsmath}
 \usepackage{babel} 
 \usepackage{txfonts}
 \usepackage{epsfig,epsf,psfrag} 
\usepackage{graphicx} 
\usepackage{pslatex}
\usepackage{epic,eepic} 
\usepackage{color,pstcol}
\usepackage{pstricks} 
\usepackage{fancyhdr} 
\usepackage{tabularx}
\usepackage{physics}
\usepackage{url}
\usepackage{listings}
\usepackage{color}
\usepackage{mathtools}
\usepackage{amssymb}
\usepackage{txfonts}
\usepackage{centernot}
\usepackage{multirow}
\usepackage[normalem]{ulem}
\useunder{\uline}{\ul}{}
\usepackage{natbib}
\usepackage{graphicx,wrapfig,subcaption,setspace,booktabs}
\usepackage{float}
\usepackage{fancyhdr}  \parindent0em  
\begin{document}
\lstset{language=Mathematica}
\title{Order from chaos in quantum walks on cyclic graphs}
\author{Abhisek Panda}
\affiliation{School of Physical Sciences, National Institute of Science Education \& Research, HBNI, Jatni-752050, India}  
\author{Colin Benjamin}
\email{colin.nano@gmail.com}
\affiliation{School of Physical Sciences, National Institute of Science Education \& Research, HBNI, Jatni-752050, India}  

\begin{abstract}
It has been shown classically that combining two chaotic random walks can yield an ordered(periodic) walk. Our aim in this paper is to find a quantum analog for this rather counter-intuitive result. We study chaotic and periodic nature of cyclic quantum walks and focus on an unique situation wherein a periodic quantum walk on $3-$cycle graph is generated via a deterministic combination of two chaotic quantum walks on the same graph. We extend our results to even numbered cyclic graphs, specifically a $4-$cycle graph too. Our results will be relevant in quantum cryptography and quantum chaos control. 
\end{abstract}
\maketitle
\section{ Introduction }
Parrando's paradox describes situations wherein a random or deterministic combination of losing strategies can yield a winning outcome. Parrondo's games seen in random walks have been shown to have significant applications in biological systems, algorithm and cryptology\cite{abbott,parrondo,tang}. Studies in recent years have shown Parrando's paradox in classical random walks wherein two chaotic walks were combined to yield an ordered  (periodic) walk\cite{cheong,buceta,kocarev,dancia,almeida}. In these papers the main focus is on combining two chaotic systems to generate order, i.e., $Chaos_{1}+Chaos_{2}=Order,$ a profound and counter-intuitive result. This result is far reaching, especially in classical chaos control theory\cite{mendoza}, etc. Our aim in this work is to find the quantum analog of this result, and we show it in the context of quantum walks on $3-$cycle graphs, see Figure 1. Quantum walks (QW's) have been  used in simulating physical systems\cite{wang} as well as in designing quantum algorithms\cite{ambainis1,kendon}. 
A QW can be described on a one dimensional lattice, or analogously on a circle with $k-$sites, with the walker starting from the origin\cite{treganna,dukes,konno}. Unlike the walker of the classical random walk, in the quantum version the walker is represented by a wave function. However, similar to a classical random walk, QW's consist of a walker and a coin. The coin in a QW, in general, is a qubit. Quantum walks can also be fashioned with coins which are qutrits\cite{jishnu2} or even qudits\cite{prl-qudit-qwalk} and also entangled coins\cite{jishnu1}. In this paper we will only focus on qubits as coins. Similar to classical random walks, in QW's if applying the coin operator on the initial coin state yields head the walker shifts to the right else the shift is to the left. In addition to head or tail, the coin state in quantum case can be in a superposition of head and tail, in which case the walker moves to a corresponding superposition of left and right lattice sites.

A QW need not be restricted to a line, QW's on a $k-$ cycle graphs have been studied in detail in Refs.~\cite{treganna,dukes}. Intriguingly, in quantum walks on $k-$cyclic graphs, chaotic behavior has been seen. A quantum walk on a $k-$cycle graph is termed periodic if it returns periodically to  a particular position, say the origin, after a finite number of steps, otherwise it's chaotic. Our aim in this work is to show that two chaotic QW's can be combined to yield a periodic walk. To fulfil our aim we proceed as follows, we first dwell on defining the shift and coin operators in a $k-$cycle graph and then find out the condition for the walker to be ordered or periodic.  Next we deal with specific coin operators which yield chaotic QW's and find out the conditions under which combination of these coins, i.e., Parrondo sequences, will generate periodic QW's. We then show via plots the probability of returning to origin in a 3-cycle graph focusing on our aim of getting periodic QW's via a Parrondo strategy of alternating between coins which yield only chaotic QW's.  Next, we show results for a $4-$cycle graph wherein similar to the $3-$cycle graph we see the combination of chaotic quantum walks leading to periodic quantum walks. Recently there have been reports on designing quantum algorithms  for quantum key generation via mixing chaotic signals\cite{nat-comms}. We at the end of this manuscript show how to do that via mixing quantum chaotic walks to generate a secure encryption-decryption mechanism. 
\begin{figure}[h!]
  \hspace{-1.3 cm} 
    \begin{subfigure}[h]{0.25\textwidth}
      \includegraphics[scale=.75]{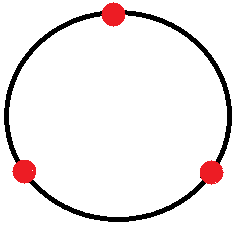} 
             \label{3-cycle}
        \end{subfigure}
    \begin{subfigure}[h]{0.23\textwidth}
          \includegraphics[scale=.95]{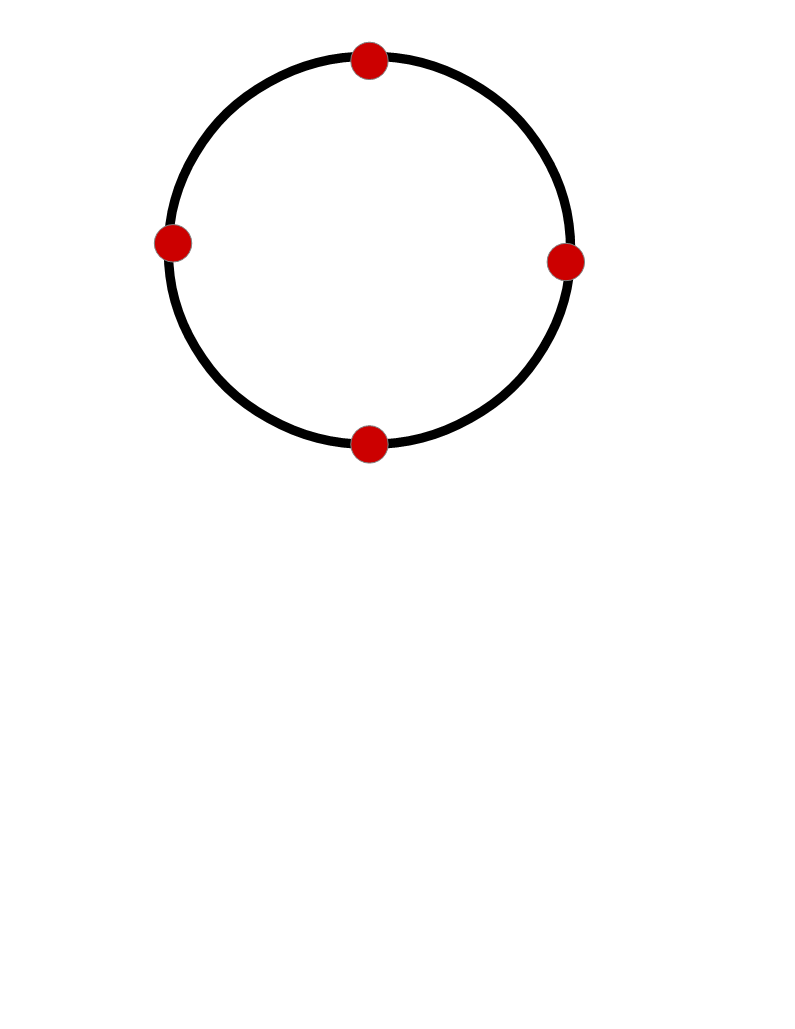}
           \label{4-cycle}
    \end{subfigure}
    \vspace{-2.5cm}
 	\caption{3-cycle graph (Bottom-Left) and 4-cycle graph (Top-Right).}
 	\label{G}
\end{figure}

\section{ Discrete time quantum walk (DTQW) on cyclic graphs} In the DTQW on a cycle, similar to that on a line, the space of walker is defined via the tensor product of position and coin space, i.e., \textbf{$H_P\otimes H_c$} where \textbf{$H_P$} is position Hilbert space and \textbf{$H_c$}  is coin Hilbert space. In case, coin is a qubit with two states $|0\rangle=\left(\begin{array}{c}1\\0\end{array}\right)$ and $|1\rangle=\left(\begin{array}{c}0\\1\end{array}\right)$, then the general unitary coin operator can be written as
\begin{equation}\begin{aligned}
 C_{2}(\rho,\alpha,\beta)=\left(\begin{array}{ll}
\sqrt{\rho} & \sqrt{1-\rho} e^{i \alpha} \\
\sqrt{1-\rho} e^{i \beta} & -\sqrt{\rho} e^{i(\alpha+\beta)}
\end{array}\right),&\\\text{ where }{}0 \leqslant \rho \leqslant 1,0 \leqslant \alpha, \beta \leqslant \pi.&
\end{aligned}\end{equation}
The walker shift to right by one site  when final state is $|1\rangle$ and to left by one site when final state is $|0\rangle$. So the shift operator for the case when walker walks on a line is given by  \begin{equation}\begin{aligned}
S =\sum_{s=0}^{1} \sum_{i=-\infty}^{\infty}|i+2 s-1\rangle\langle i|\otimes| s\rangle\langle s|,
\end{aligned}\end{equation}
but to perform QW on a circle of '$k$' sites the shift operator has to be modified\cite{dukes} as shown below
\begin{equation}\begin{array}{l}
S=\sum_{s=0}^{1} \sum_{i=0}^{k-1}|(i+2s-1)\mod k\rangle\langle i|\otimes| s\rangle\langle s|.
\end{array}
\label{shift}
\end{equation}
Using Eq.~(\ref{shift}) and the coin operator we can represent the QW by an unitary operator as
\begin{equation}
U_{k}=S \cdot\left(I_{k} \otimes C_{2}\right).
\end{equation}
The matrix $U_k$ is a $2\times2$ block circulant matrix, see also Refs.~\cite{dukes,treganna}. It is represented as, \begin{equation}\begin{array}{c}
U_{k}=\operatorname{CIRC}_{k}\left(\left[\begin{array}{cc}
0 & 0 \\
0 & 0
\end{array}\right]_{0},\left[\begin{array}{cc}
\sqrt{\rho} & \sqrt{1-\rho} e^{i \alpha} \\
0 & 0
\end{array}\right]_{1},\right.\\\left.\left[\begin{array}{cc}
0 & 0 \\
0 & 0
\end{array}\right]_{2}, \cdots, 
\left[\begin{array}{cc}
0 & 0 \\
\sqrt{1-\rho} e^{i \beta} & -\sqrt{\rho} e^{i(\alpha+\beta)}
\end{array}\right]_{k-1}
\right).
\end{array}\end{equation}
The walker repeatedly applies $U_k$ to its initial state to get to its final state. For $N$ steps we get, $U_{k}^{N}\left|\psi_{i}\right\rangle=\left|\psi_{f}\right\rangle$. If the walker return to its initial state after $N$ steps for any arbitrary initial quantum state then the QW by the walker is said to be periodic, hence for periodic QW we have \begin{equation}
 U_{k}^{N}\left|\psi_{i}\right\rangle=\left|\psi_{i}\right\rangle.
 \label{ukn}
\end{equation} 
Now let eigen-vectors of $U_k$ be  $\{|x_i\rangle\}$ and the corresponding eigenvalue be $\{\lambda_i\}$, as $|\psi_i\rangle$ is arbitrary we can represent it in terms of eigen-vectors of $U_k$ as $\left|\psi_{i}\right\rangle=\sum_{j=1}^{2 k} \alpha_{j} |x_{j}\rangle$. Applying $U_k$,  $N$ times on an initial state we get \begin{equation}
U_{k}^{N}\left|\psi_{i}\right\rangle=\sum_{j=1}^{2 k} \alpha_{j} \lambda_{j}^{N} |\mathbf{x}_{j}\rangle.
\label{lambda}
\end{equation}
Comparing Eq.~(\ref{lambda}) with Eq.~(\ref{ukn}) we get the the condition of periodicity as
\begin{equation}
\begin{aligned}
&\lambda_{j}^{N}=1\text{ , } \forall\text{ } 1\leq j \leq 2k, \text{ or }
U_{k}^{N}=I_{2 k}.
\end{aligned}
\label{period}
\end{equation}
In case a particular unitary operator satisfies Eq.~\ref{period} then it gives periodic result and that operator is said to generate a periodic or ordered QW, while if the QW does not satisfy Eq.~(\ref{period}) it is called chaotic.

\subsection{ Block diagonalizing $U_k$}
By diagonalizing $U_k$ we will get it's eigenvalues which then simplifies the problem of finding periodicity of the QW on cyclic graphs. As $U_k$ is a circulant matrix it is block diagonalized by a tool known as commensurate Fourier matrix as was also done in Ref.~\cite{dukes}. The commensurate Fourier matrix of $M$ dimension is defined as $$F^{M}=\left(F_{m, n}^{M}\right)=\frac{1}{\sqrt{M}}\left(e^{2 \pi i \frac{m n}{M}}\right).$$ Now as our matrix is of dimension $k\otimes2$ we define $F=F^{k\otimes2}$ and $U_{k'}=FU_kF^{\dagger}$. The matrix $U_{k'}$ has the form \begin{equation}
    U_{k'} =\left(\begin{array}{cccc}
U_{k, 0} & 0 & \dots & 0 \\
0 & U_{k, 1} & \dots & 0 \\
\vdots & \vdots & \ddots & \vdots \\
0 & 0 & \dots & U_{k, k-1}
\end{array}\right) , \label{ukl}
\end{equation} where each $U_{k,l}$ is a block $2\cross2$ matrix.
The eigenvalues of each such block $U_{k,l}$ is given by
\begin{equation}\begin{aligned}
\lambda_{k, l}^{\pm}=&\frac{1}{2} e^{-2 \pi i \frac{l}{k}}\left(\left(1-e^{4 \pi i\frac{l}{k}+i \delta}\right) \sqrt{\rho}\right.\\&\left.\pm2 \sqrt{e^{4 \pi i\frac{l}{k}+ i \delta}\left(1-\rho \sin ^{2}\left[\frac{2 \pi l}{k}+\frac{\delta}{2}\right]\right)}\right).
\end{aligned}
\label{eigenvalue}
\end{equation} 
Now Eq.~(\ref{period}) is satisfied if both eigenvalues, i.e, $\lambda_{k,l}^{\pm}$ take the form of a de Movire number, i.e., each $\lambda_{j}=e^{2\pi i \frac{m_j}{n_j}}$ which can be equivalently written as
\begin{equation}\lambda_{k, l}^{\pm}=e^{2 \pi i \frac{m_{j}}{n_{j}}} \text{ or  }\lambda_{k, l}=\frac{\lambda_{k, l}^{+}+\lambda_{k, l}^{-}}{2}=e^{2 \pi i \frac{m_{j}}{n_{j}}},
\label{demoivre}
\end{equation}
and the least common multiple of $\{n_j\}$=N, with  $1 \leq j\leq 2k$, this gives periodicity of QW on $k-$cyclic graph to be $N$. Examples of periodic and chaotic QW's can be seen in Refs.~\cite{treganna,dukes} which satisfy Eqs.~(\ref{eigenvalue},\ref{demoivre}) and are also given below in Figure~2 for parameters mentioned in the figure caption.
\begin{figure}[h]
    \centering
\includegraphics[width=9cm]{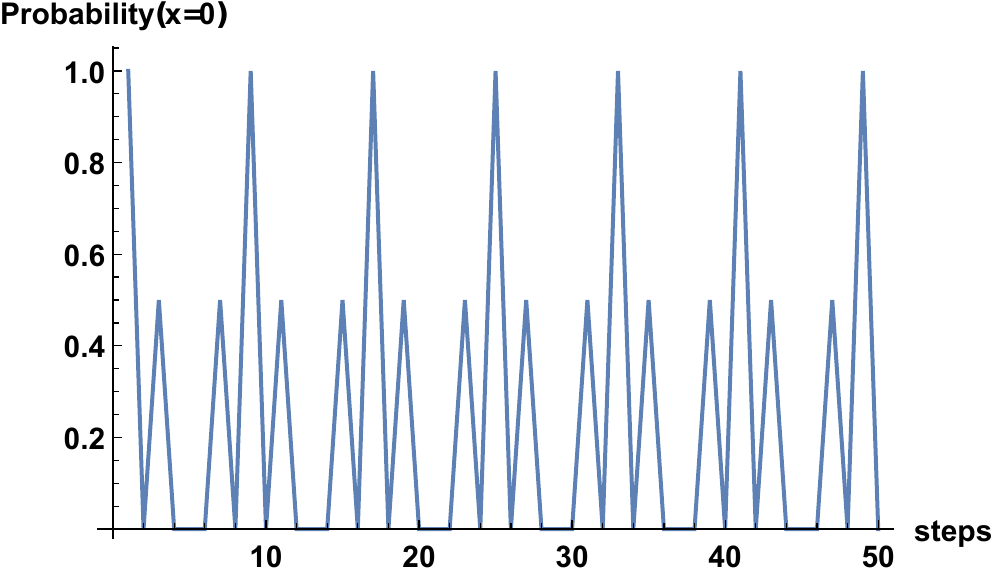}
\includegraphics[width=9cm]{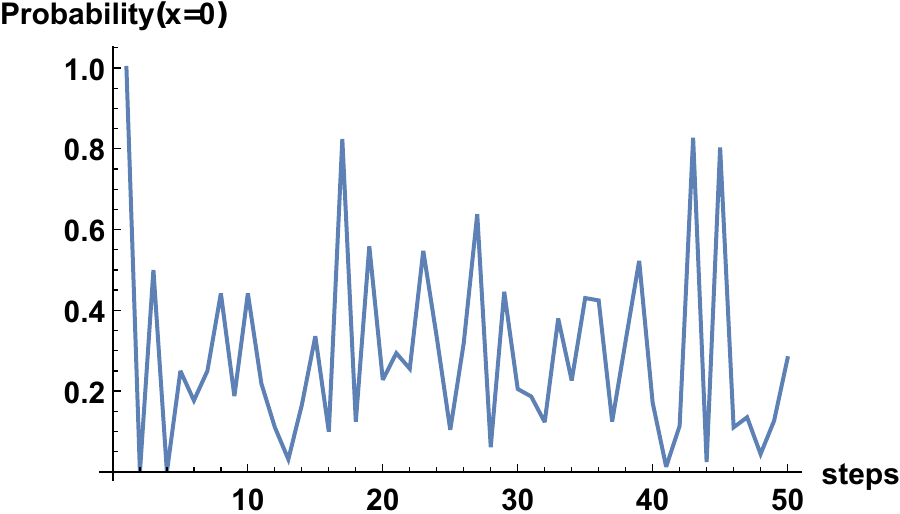}
    \caption{Probability for finding the walker at its initial site at $x = 0$ for $4-$cycle graph (upper, ordered, i.e., periodic with period 8) and $5-$cycle graph (lower, chaotic). Probability is plotted against time steps for the QW with a Hadamard coin $H=C_{2}(\frac{1}{2},0,0)$. Only even time steps are plotted.}
    \label{fig:my_label}
\end{figure}
\section{ Parrondo strategies ($ABAB..$) in DTQW on $3-$cycle graphs} In the subsequent analysis in this section we stick to only QW's on $3-$cycle graphs, see Fig.~1 (left bottom). We consider three different unitary operators $A=U_{3}(\rho_1,\alpha_1,\beta_1)=S\cdot (I_{3}\times C_{2}(\rho_1,\alpha_1,\beta_1))$, $B=U_{3}(\rho_2,\alpha_2,\beta_2)=S\cdot (I_{3}\times C_{2}(\rho_2,\alpha_2,\beta_2)$ and $C=U_{3}(\rho,\alpha,\beta)=S\cdot (I_{3}\times C_{2}(\rho,\alpha,\beta)$, with coin operator $C_{2}$ defined in Eq.~(2) and $U_{3}$ in Eq.~(4) with $k=3$. We note that only QW obtained from unitary operator $C$ satisfies Eq.~(\ref{ukn}) and gives ordered QW with period $N$, while unitary operators $A,B$ lead to chaotic QW's. Our aim is to find a suitable combination of $A,B$ coin operators which would give an ordered (periodic) QW. To achieve our goal, we first check the Parrondo sequence $ABAB...$ of unitary operators and calculate the eigenvalues of the matrix $FABF^{\dagger}$. For the $3-$cycle graph, there will be $3$ diagonal blocks $U_{3,0}, U_{3,1}$ and $U_{3,2}$ see Eq.~(\ref{ukl}). The sum of eigenvalues, for the diagonal block: $U_{3,1}$ are \begin{eqnarray}\lambda_{3,1}^{AB}=&\frac{ \lambda_{3,1 }^{AB + }+\lambda_{3,1}^{AB-}}{2}=
(2 e^{i \alpha_2+i \beta_1} \sqrt{1-\rho_1} \sqrt{1-\rho_2}\nonumber\\&+2 e^{i \alpha_1+i \beta_2} \sqrt{1-\rho_1} \sqrt{1-\rho_2}+2(-1)^{2 / 3} \sqrt{\rho_1} \sqrt{\rho_2}\nonumber\\&-2(-1)^{1 / 3} e^{i \alpha_1+i \alpha_2+i \beta_1+i \beta_2} \sqrt{\rho_1} \sqrt{\rho_2})/4.
\end{eqnarray}

Since our aim is to check if the Parrondo combination $ABAB...$ leads to periodicity. We now evaluate the eigenvalues of matrix [$FCCF^{\dagger}$], remembering that unitary operator $C$ generates a periodic QW. The eigenvalues, for a $3-$cycle graph with coin $CC..$, and for the same diagonal block: $U_{3,1}$ is given as  \begin{eqnarray}\lambda_{3,1}^{CC}=\frac{ \lambda_{3,1 }^{CC + }+\lambda_{3,1}^{CC-}}{2}=&
(4 e^{i( \alpha+\beta)}+2(-1)^{2 / 3}\rho-4 e^{i(\alpha +\beta)}\rho\nonumber\\
&-2(-1)^{1 / 3}e^{2 i (\alpha+\beta)}\rho)/4.\end{eqnarray}
\begin{figure}[h]
        \centering
    \includegraphics[width=9cm]{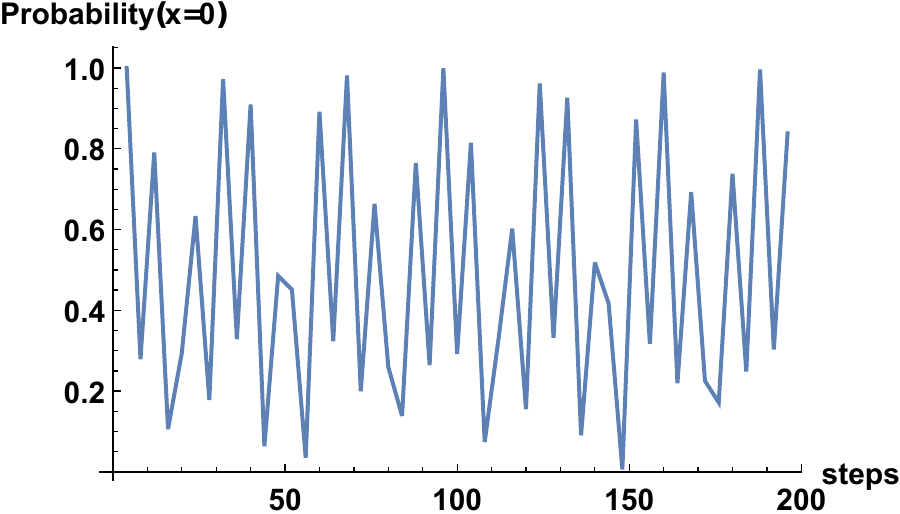} \caption{QW on $3-$cycle graph by repeatedly applying unitary $A=U_{3}(\rho_{1},0,0)$, which results in chaotic quantum walk.}%
    \label{fig:example}
    \end{figure}
    \begin{figure}[h]
        \centering
    \includegraphics[width=9cm]{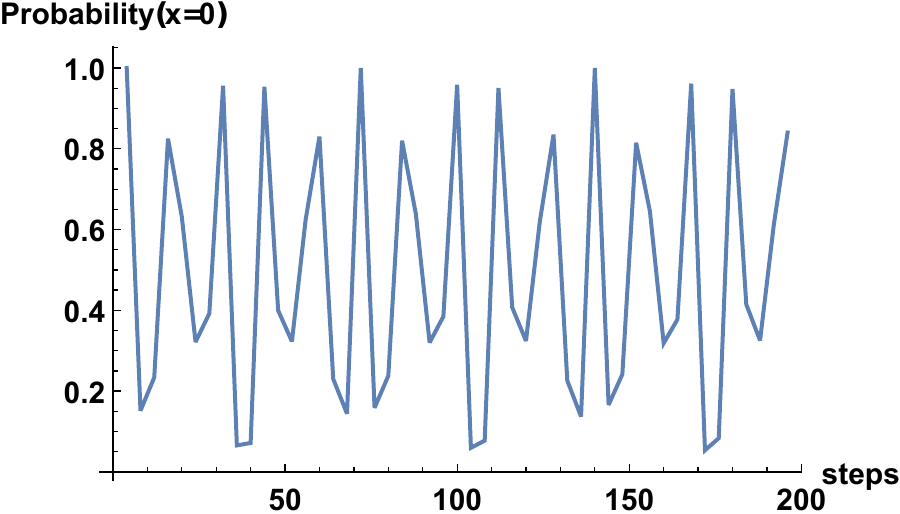} \caption{Chaotic QW on 3-cycle graph obtained by repeatedly applying unitary $B=U_{3}(\rho_{2},0,0)$.}%
    \label{fig:example}
    \end{figure} 
    \begin{figure}[h]
        \centering
    \includegraphics[width=9cm]{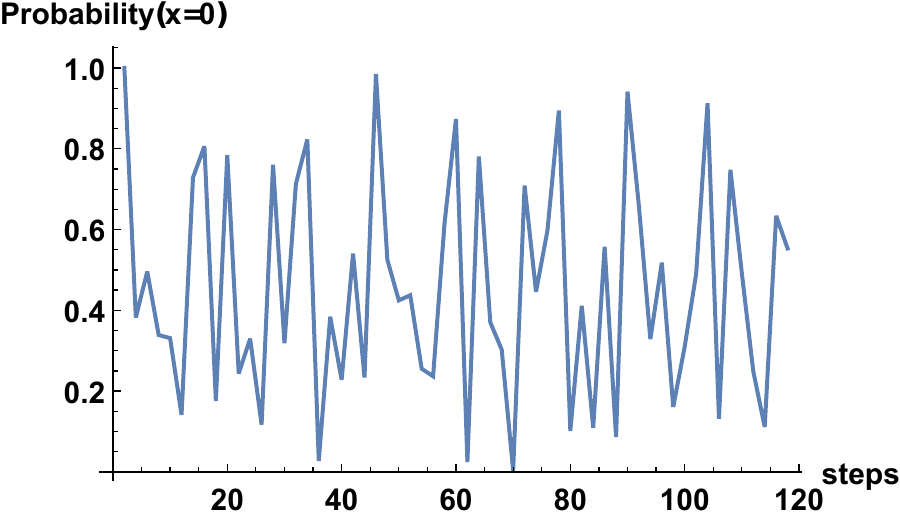} \caption{Chaotic QW on $3-$cycle graph by repeatedly applying the Parrondo  sequence $ABAB..$. Plotting every second point.}
    \label{fig:example}
    \end{figure}
Since repeatedly applying $C$ generates a periodic QW, so if repeatedly applying $AB$ were to also generate a periodic QW, then the form of the eigenvalues should match, i.e., \begin{equation}\lambda_{3,1}^{CC}=\lambda_{3,1}^{AB}.
\label{ev:ab=c}\end{equation}
 Repeating above mentioned procedure for other diagonal blocks $U_{3,0}$ and $U_{3,2}$, we get exactly similar equations to that shown in Eq.~(\ref{ev:ab=c}).
 Taking $\alpha_1=\alpha_2=\alpha$ and $\beta_1=\beta_2=\beta$ and
equating the coefficients of frequencies on both sides we get two equations as follows 
\begin{equation}
\rho_1 \cdot \rho_2=\rho^2  \mbox{,  and   } \sqrt{1-\rho_1} \sqrt{1-\rho_2}=1-\rho.
\label{AB-condition}
\end{equation}
The only solution to Eq.~(\ref{AB-condition}) is $\rho_1=\rho_2=\rho$, which gives the trivial solution $A=B$, as $\alpha_1=\alpha_2=\alpha$ and $\beta_1=\beta_2=\beta$. This solution is not desirable as the deterministic combination of $A$ and $B$ gives again a chaotic QW. Following this, similar calculations done for deterministic combinations $ABB, ABBB, ABBBB$ and $AABB$ do not  generate ordered(periodic) QW's. The solutions for $ABBB$ and $ABBBB$ were trivial as in case of $AB$. However, $ABB$ gives a non-trivial solution for a small range of $\rho$, one could find suitable coin operators $A$ and $B$ for which $ABB$ is periodic but unfortunately we weren't able to zero in on any exact values for $\rho_{1}, \rho_{2}$ and $\rho$. Nonetheless, the deterministic combination $AABB...$ does generates an ordered(periodic) QW as shown below.
\section{\bf \em Parrondo strategies ($AABB..$) in DTQW on $3-$cycle graphs}   We start by calculating the eigenvalues of the matrix $FAABBF^{\dagger}$. Similar to the case preceding, for the $3-$cycle graph, there will be $3$ diagonal blocks $U_{3,0}, U_{3,1}$ and $U_{3,2}$ see Eq.~(\ref{ukl}). The sum of eigenvalues, for the diagonal block: $U_{3,1}$  are 
\begin{eqnarray}
&\lambda_{3,1}^{AABB}=\frac{\lambda_{3,1 }^{AABB+}+\lambda_{3,1}^{AABB-}}{2}=4 e^{i (\alpha_1+ \alpha_2+ \beta_1+ \beta_2)}\nonumber\\
&+2(-1)^{2 / 3} e^{i (\alpha_2+ \beta_2)} \rho_1-4 e^{i \alpha_1+i \alpha_2+i \beta_1+i \beta_2} \rho_1\nonumber\\
&-2(-1)^{1/3} e^{2 i \alpha_1+i \alpha_2+2 i \beta_1+i \beta_2} \rho_1+2(-1)^{2 / 3} e^{i \alpha_1+i \beta_1} \rho_2\nonumber\\
&-4 e^{i \alpha_1+i \alpha_2+i \beta_1+i \beta_2} \rho_2-2(-1)^{1 / 3} e^{i \alpha_1+2 i \alpha_2+i \beta_1+2 i \beta_2} \rho_2\nonumber\\
&+ 2(-(-1)^{1 / 3}-(-1)^{2 / 3} e^{i(\alpha_1+\beta_1)}-(-1)^{2 / 3} e^{i(\alpha_2+\beta_2)}\nonumber\\
&+2 e^{i(\alpha_1+\alpha_2+\beta_1+\beta_2)}+(-1)^{2 / 3} e^{2 i(\alpha_1+\alpha_2+\beta_1+\beta_2)}\nonumber\\
&+(-1)^{1 / 3} e^{i (2 \alpha_1+\alpha_2+2 \beta_1+\beta_2)}
+(-1)^{1 / 3} e^{i(\alpha_1+2 \alpha_2+\beta_1+2 \beta_2)}) \rho_1 \rho_2\nonumber\\
&-2(e^{i(\alpha_2+\beta_1)}+e^{i(\alpha_1+\beta_2)})(-(-1)^{2/3}+e^{i(\alpha_1+\beta_1)}+e^{i(\alpha_2+\beta_2)}\nonumber\\
&+(-1)^{1 / 3} e^{i(\alpha_1+\alpha_2+\beta_1+\beta_2)}) \sqrt{(1-\rho_1) \rho_1 (1-\rho_2) \rho_2}.\end{eqnarray}  Since our aim is to check if the Parrondo combination $AABB...$ leads to periodicity. We now evaluate the eigenvalues of matrix [$FCCCCF^{\dagger}$], remembering that coin operator $C$ generates a periodic QW. The eigenvalues, for a $3-$cycle graph with coin operators $CCCC..$ being applied successively and for the same diagonal block $U_{3,1}$ is given as 
\begin{eqnarray}
&\lambda_{3,1}^{CCCC}=\frac{\lambda_{3,1 }^{CCCC+ }+\lambda_{3,1}^{CCCC-}}{2}=4e^{2i(\alpha+\beta)}+8(-1)^{2/3}e^{i(\alpha+\beta)}\rho\nonumber\\
&-16e^{2i(\alpha+\beta)}\rho-8(-1)^{1/3}e^{3i(\alpha+\beta)}\rho-2(-1)^{1/3}\rho^{2}\nonumber\\
&-8(-1)^{2/3}e^{i(\alpha+\beta)}\rho^{2}+12e^{2i(\alpha+\beta)}\rho^{2}\nonumber\\&+8(-1)^{1/3}e^{3i(\alpha+\beta)}\rho^{2}+2(-1)^{2/3}e^{4i(\alpha+\beta)}\rho^{2}.\end{eqnarray}
Since repeatedly applying $C$ gives a periodic QW, so if repeatedly applying $AABB$ were to generate a periodic QW, then the form of the eigenvalues should match, i.e.,
\begin{equation}\lambda_{3,1}^{CCCC}=\lambda_{3,1}^{AABB}.\label{ev:aabb=c}\end{equation}
Similar equations as has been shown in Eq.~\ref{ev:aabb=c} can be written for other  blocks, i.e., $U_{3,0}$ and $U_{3,2}$. Taking $\alpha_1=\alpha_2=\alpha$ and $\beta_1=\beta_2=\beta$ and matching frequencies both sides we get\begin{eqnarray}
\rho_1+  \rho_2-2  \rho_1  \rho_2+2 \sqrt{(1-\rho_1) (1-\rho_2) \rho_1\rho_2}&=4   \rho-4\rho^2,\nonumber\\
\rho_1+ \rho_2- \rho_1 \rho_2+2 \sqrt{(1-\rho_1) (1-\rho_2) \rho_1\rho_2}&=4 \rho-3 \rho^{2},\nonumber\\
\rho_1 \cdot \rho_2  &= \rho^2 \mbox{.  \hskip .8 cm     }
\label{aabb=cc}\end{eqnarray}
From above three equations only two are independent. The third can be derived from other two. The solution to Eq.~(\ref{aabb=cc}) is given by 
\begin{equation}\begin{aligned}
 &\rho_1 \rightarrow 3 \rho-4 \rho^{2}\pm2 \sqrt{2} \sqrt{\rho^{2}\left(1-3 \rho+2 \rho^{2}\right)}.\\& \rho_2 \rightarrow 3 \rho-4 \rho^{2}\mp2 \sqrt{2} \sqrt{\rho^{2}\left(1-3 \rho+2 \rho^{2}\right)}.
\end{aligned}
\label{eq:aabb=c}
\end{equation}
This gives the possibility of $A$ and $B$ being chaotic but $AABB$ being periodic.
\begin{figure}[h]
        \centering
    \includegraphics[width=9cm]{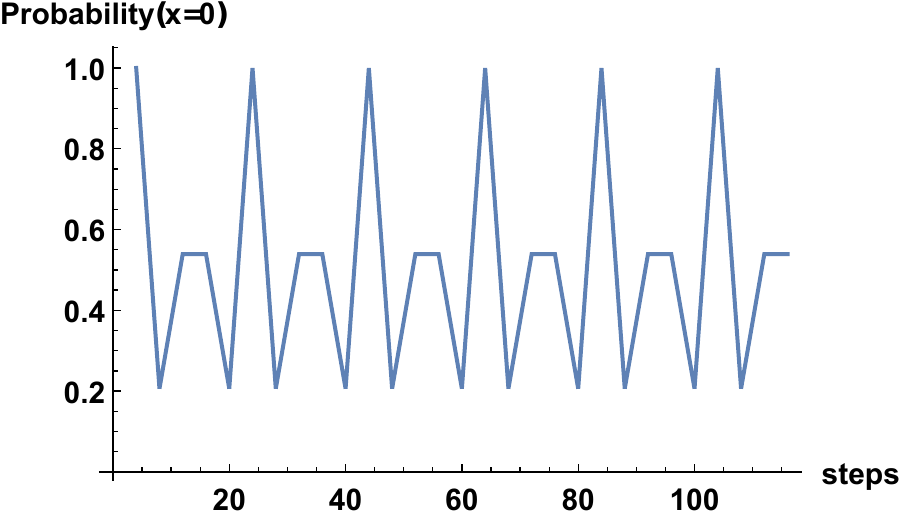} \caption{ Ordered QW on $3-$cycle graph by repeatedly applying Parrondo sequence $AABB...$. Every fourth point is plotted. Quantum walk is periodic with periodicity $20$.}%
    \label{fig:example}%
    \end{figure} 
    \begin{figure}[h]
        \centering
    \includegraphics[width=9cm]{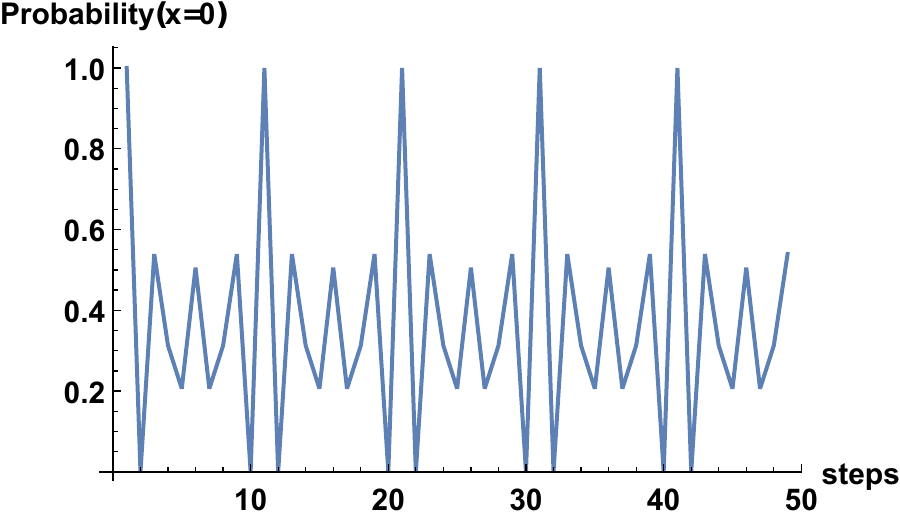} \caption{ Ordered QW on $3-$cycle graph by repeatedly applying unitary $C$. Every fourth point is plotted. Quantum walk is periodic with periodicity $10$.}
    \label{fig:example}%
    \end{figure}   
    
    \section{\bf \em Parrondo strategies ($A'A'B'B'..$) in DTQW on $4-$cycle graphs}
    Similar to $3-$cycle graph, for the $4-$cycle graph we consider  three different unitary operators $A'=U_{4}(\rho_{41},\alpha_1,\beta_1)=S\cdot (I_{4}\times C_{2}(\rho_{41},\alpha_1,\beta_1))$, $B'=U_{4}(\rho_{42},\alpha_2,\beta_2)=S\cdot (I_{4}\times C_{2}(\rho_{42},\alpha_2,\beta_2)$ and $C'=U_{4}(\rho_{4},\alpha,\beta)=S\cdot (I_{4}\times C_{2}(\rho_{4},\alpha,\beta)$, with coin operator $C_{2}$ defined in Eq.~(2) and $U_{4}$ in Eq.~(4) with $k=4$. We note that only QW obtained from unitary operator $C'$ satisfies Eq.~(\ref{ukn}) and gives ordered QW with period $N$, while unitary operators $A', B'$ lead to chaotic QW's. We check the Parrondo sequence $A'A'B'B'...$ of unitary operators and calculate the eigenvalues of the matrix $FAABBF^{\dagger}$. For the $4-$cycle graph, there will be $4$ diagonal blocks $U_{4,0}, U_{4,1}, U_{4,2}$ and $U_{4,3}$ see Eq.~(\ref{ukl}). Following exactly the same procedure as was adopted for $3-$cycle walks, one gets an identical set of equations Eq.~(\ref{aabb=cc}-\ref{eq:aabb=c}), from which we get the condition for unitaries $A'$ and $B'$ generating chaotic quantum walks in $4-$cycle graph but $A'A'B'B'$ generating a periodic quantum walk. In the Results subsection on 4-cycle graph, we give the details of parameters which lead to desired outcome of two chaotic quantum walks $A', B'$ combining in sequence $A'A'B'B'$ to generate a periodic quantum walk in a $4-$cycle graph. 
   \begin{figure}[h]
         \includegraphics[width=9cm]{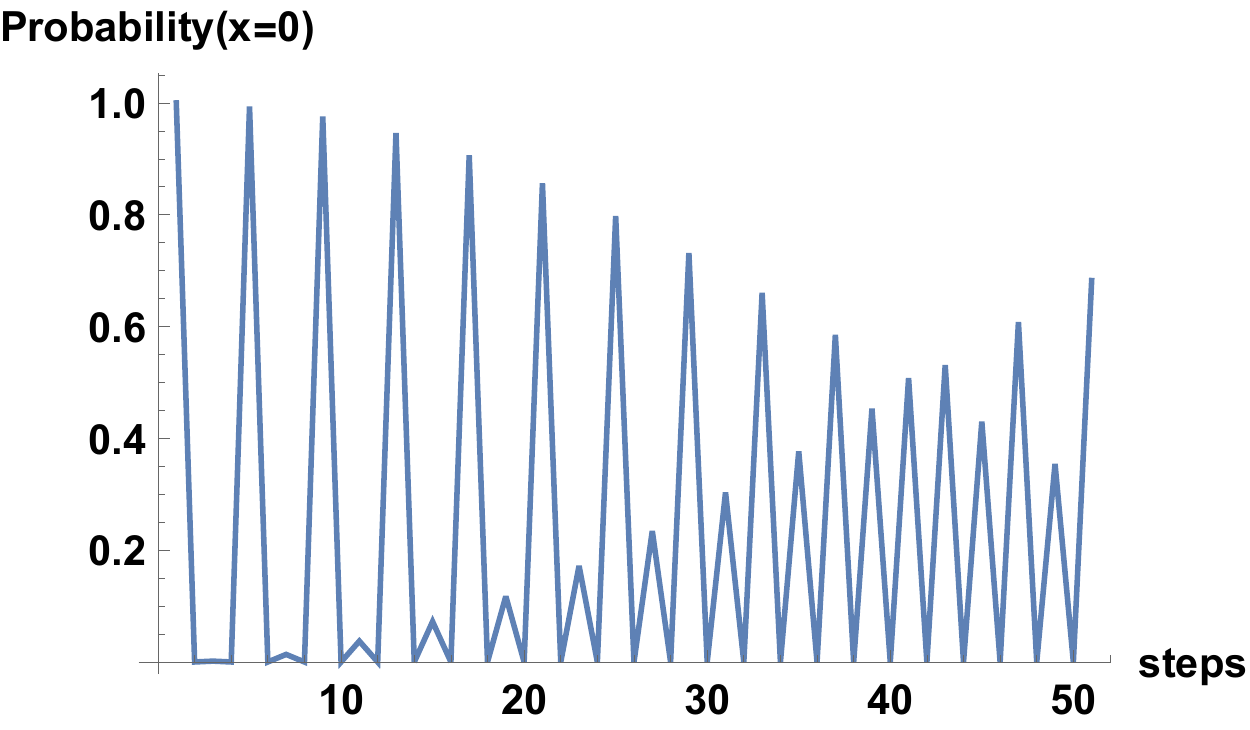} \caption{QW on $4-$cycle graph by repeatedly applying unitary $A'=U_{4}(\rho_{41},0,0)$, which results in chaotic quantum walk.}%
    \label{fig:example}
    \end{figure}
    \begin{figure}[h]
        \centering
    \includegraphics[width=9cm]{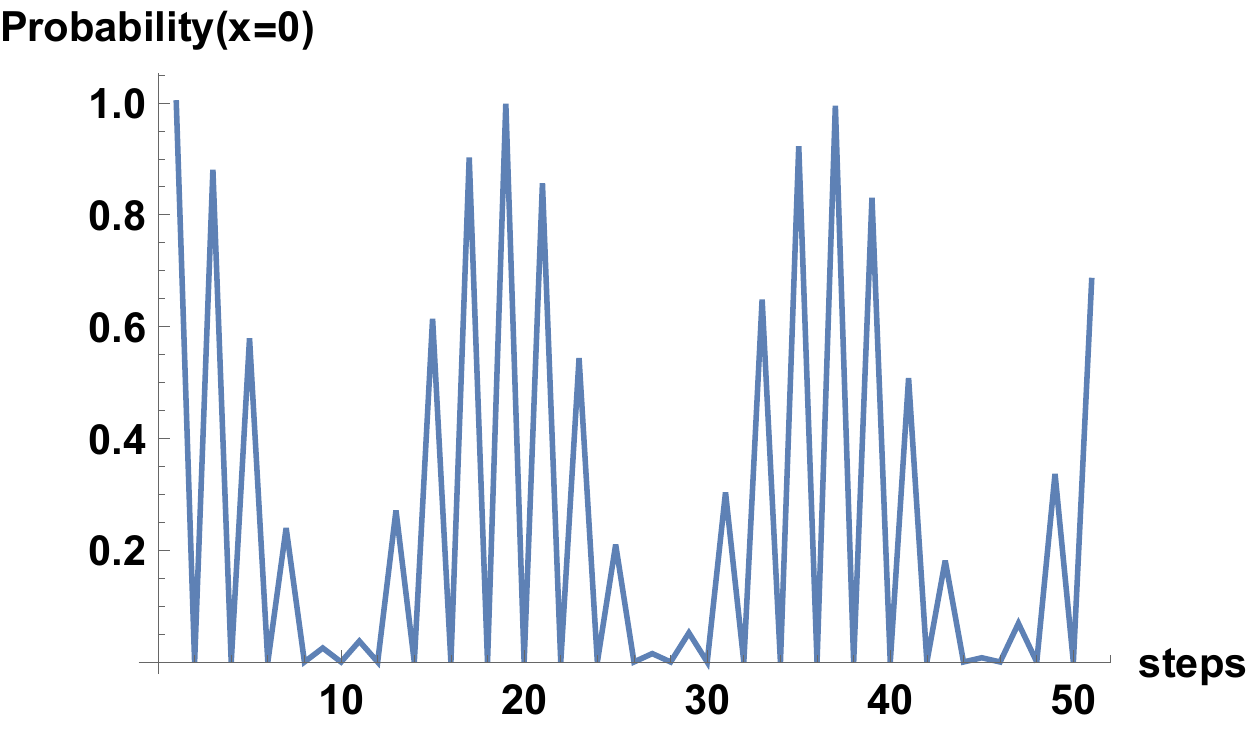} \caption{Chaotic QW on $4-$cycle graph obtained by repeatedly applying unitary $B'=U_{4}(\rho_{42},0,0)$.}%
    \label{fig:example}
    \end{figure}   \begin{figure}[h]
        \centering
    \includegraphics[width=9cm]{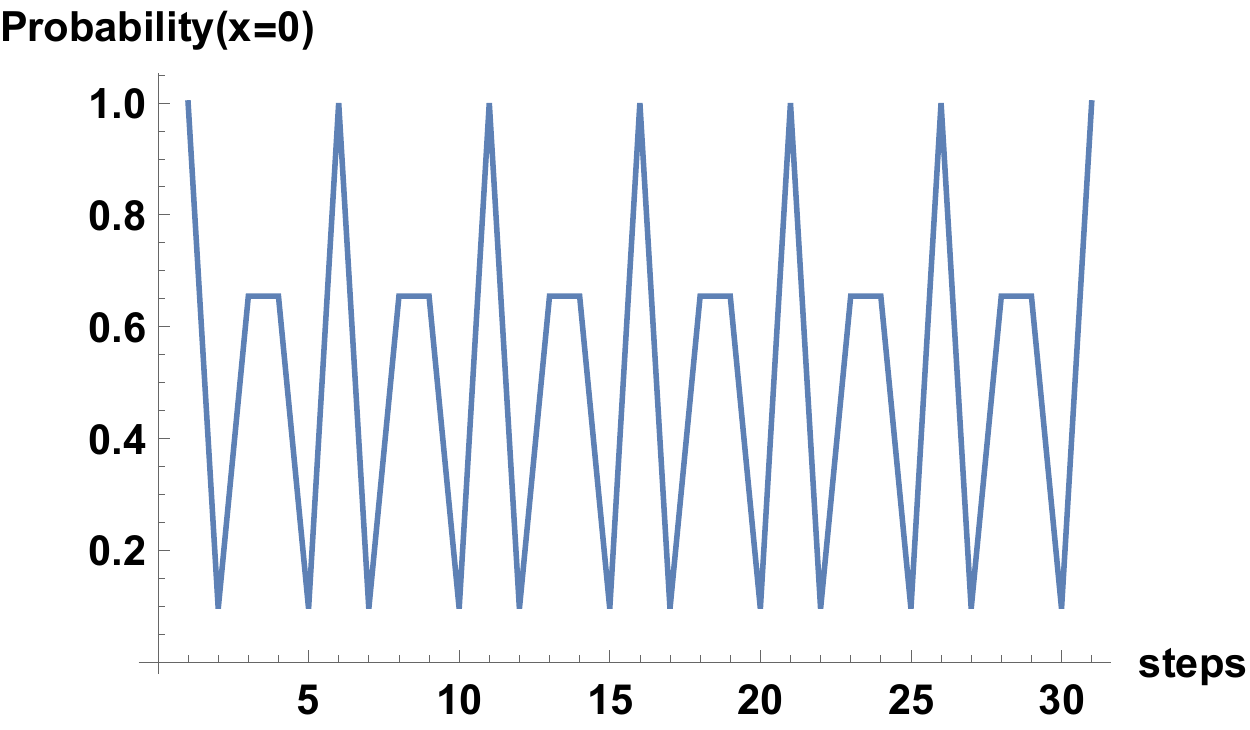} \caption{ Ordered QW on $4-$cycle graph by repeatedly applying Parrondo sequence $A'A'B'B'...$. Quantum walk is periodic with periodicity $5$.}%
    \label{fig:example}%
    \end{figure} 
    \begin{figure}[h]
        \centering
    \includegraphics[width=9cm]{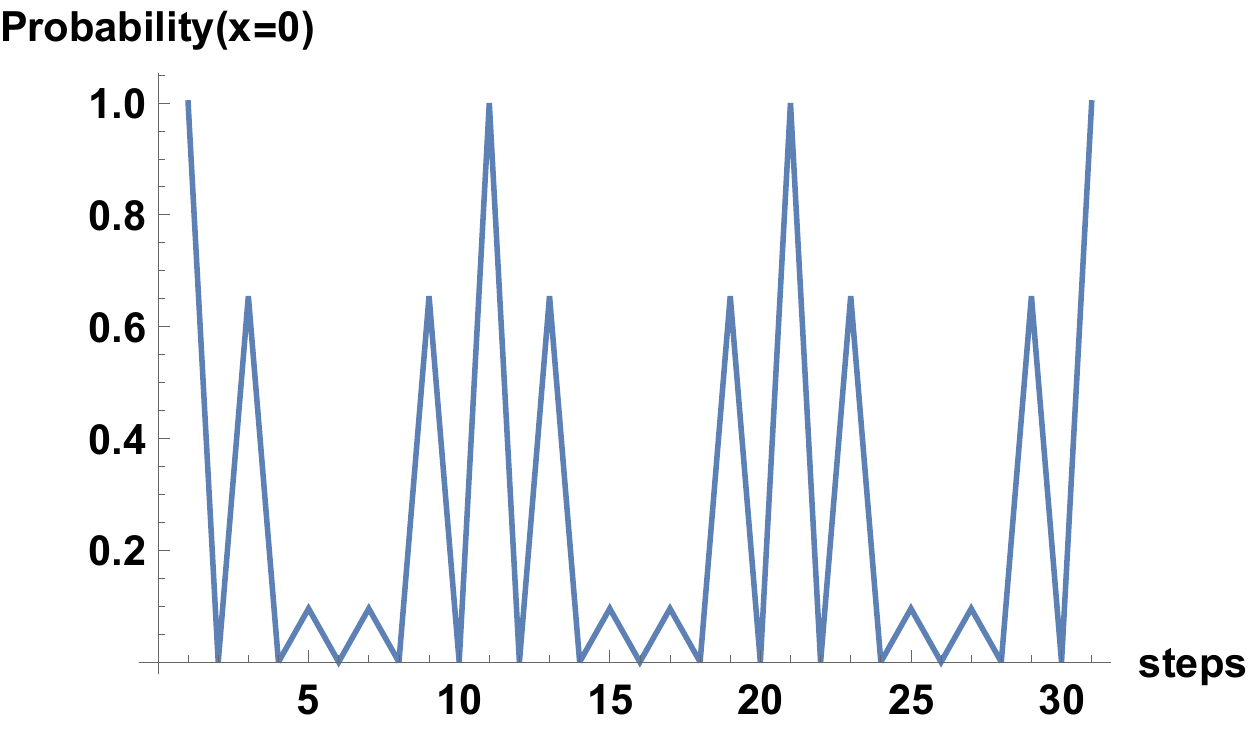} \caption{ Ordered QW on $4-$cycle graph by repeatedly applying unitary $C'$. Quantum walk is periodic with periodicity $10$.}
    \label{fig:example}%
    \end{figure}   
\section{ Results}
\subsection{$3-$cycle graph} 
In Refs.\cite{treganna,dukes}, table of examples have been given which satisfy Eq.~(\ref{ukn}). From these examples, and considering $k=3$ with  $\rho=\frac{(5-\sqrt{5})}{6}=0.460655$, $\alpha=0$ and $\beta=0$ and using Eq.~(\ref{eq:aabb=c}) with $\alpha_1=\alpha_2=\alpha=0$, $\beta_1=\beta_2=\beta=0$ we get 
\begin{equation}
\rho_{1}=0.264734,\mbox{    } \rho_{2}=0.801571.
\end{equation}
Considering unitary operators $A=U_{3}(\rho_1,0,0)$, and $B=U_{3}(\rho_2,0,0)$ while $C=U_{3}(\rho,0,0)$ and repeating the calculations done in Eqs.~(\ref{ev:ab=c},\ref{AB-condition}) shows that one cannot get periodicity, i.e., coins $ABABAB..$ generate a  chaotic QW when quantum walks resulting from repeated application of coins $AAA...$ and $BBB...$ are chaotic too. This is shown in Figs.~ 3, 4 and 5. The initial state of the walker is taken as $|1\rangle\otimes|0\rangle$ and the probability of walker to be in initial site $0$ versus number of steps of walker is plotted.
   In Figure~6, we show that the combination of unitary operators $AABB$ does generate an ordered (periodic) QW fulfilling our aim to show that combining two chaotic systems can in certain situations lead to an ordered or periodic outcome. The periodicity of the  combination $AABB$ is $20$. It is to be noted that the combination $CCCC$ plotted in Figure~7 gives periodicity $10$. The reason for this is because the value of every fourth point in figure 6 is equal to the value of every fourth point in figure 7, i.e., probability for finding the walker at $x=0$, at time steps $4$, $8$, and $12$, etc is same, due to which we miss periodicity of $10$ and get  periodicity  of $20$ for $AABB..$. Another method of determining whether the walk is periodic or chaotic is via calculating the Lyapunov exponent. In Appendix, section~\ref{app}, we give a short recipe for calculating this. For chaotic quatum walks the Lyapunov exponent\cite{lyapunov} is positive while for periodic walks it vanishes. We indeed verify these results for the unitary operators $AAAA..$ and $BBBB..$ which give a finite positive value for the Lyapunov exponent and which generate chaotic walks. In case of unitary operators $AABB..$ we get a vanishing Lyapunov exponent which confirms a periodic quantum walk.
   \subsection{$4-$cycle graph}
   Although we have shown results only in case of  $3-$ cycle graph, our work can be easily generalized to any arbitrary $k-$ cycle graph. In fact the equations obtained for a $4-$ cycle graph are identical to that shown in Eqs.~(12-20). This is no doubt a consequence of the properties of the cyclic graph itself as well as that of the commensurate Fourier matrix $F$.
 In Refs.\cite{treganna,dukes}, table of examples have been given which satisfy Eq.~(\ref{ukn}). From these examples, and considering $k=4$ with  $\rho_{4}=\frac{(5-\sqrt{5})}{8}=0.345492$, $\alpha=0$ and $\beta=0$ and using Eq.~(\ref{eq:aabb=c}) with $\alpha_1=\alpha_2=\alpha=0$, $\beta_1=\beta_2=\beta=0$ we get 
\begin{equation}
\rho_{41}=0.998489,\mbox{    } \rho_{42}=0.119545.
\end{equation}
Considering unitary operators $A'=U_{4}(\rho_{41},0,0)$, and $B'=U_{4}(\rho_{42},0,0)$ while $C'=U_{4}(\rho_{4},0,0)$ and repeating the calculations done in Eqs.~(\ref{ev:ab=c},\ref{AB-condition}) shows that quantum walks resulting from repeated application of coins $A'A'A'...$ and $B'B'B'...$ are chaotic. This is shown in Figs.~ 8, 9. The initial state of the walker is taken as $|1\rangle\otimes|0\rangle$ and the probability of walker to be in initial site $0$ versus number of steps of walker is plotted.  In Figure~10, we show that the combination of unitary operators $A'A'B'B'$ does generate an ordered (periodic) QW fulfilling our aim to show that combining two chaotic systems can in certain situations lead to an ordered or periodic outcome. The periodicity of the  combination $A'A'B'B'$ is $5$. It is to be noted that the combination $C'C'C'C'$ plotted in Figure~11 gives periodicity $10$.

  \section{Secure encryption-decryption mechanism via mixing chaotic quantum walks} In a recent work\cite{intjqi}, quantum cryptographic protocol based on a quantum walk has been proposed. We tweak this proposal by implementing it with chaotic quantum walks. This adds a further  layer of security on top of that shown in Ref.~\cite{intjqi}. These are the steps. (i) {\underline {generating public key}}- If Bob has to send a message $m\in \{0, 1, 2\}$ to Alice, then Alice will make a public key using the unitary operator $B$ and the state of the walker $|l\rangle|s\rangle$ as follows $|\psi_{PK}\rangle = BB|l\rangle|s\rangle$. Here, $B$ is the unitary operator that generates the chaotic quantum walk as shown in Fig.~4, $|\psi_{PK}\rangle$ is a public key.$|l\rangle$ is the walker state on a cyclic graph while $|s\rangle$  is coin state. We also note that another coin unitary $A$ can generate a chaotic quantum walk too, see Fig.~3. Further, as we have shown in Fig.~6, the Parrondo sequence $AABB$ generates a periodic chaotic quantum walk with periodicity $20$ meaning $(AABB)^{5}=I$, with $I$ being identity. Alice after generating this chaotic state $|\psi_{PK}\rangle$ which acts as the public key sends it to Bob. (ii) {\underline {encrypting the message}}- After Bob receives the public key, he encodes message $m$ as follows- $|\psi(m)\rangle =(T_{m} \otimes I_{c})|\psi_{PK}\rangle$ where $T_{m} =\sum_{i=0}^{N-1} |i + m  \mod 3 \rangle\langle i|$ is akin to the shift operator used in Eq.~(3) for position state, while $I_{c}$ is the identity operator acting on coin state. (iii) {\underline {decrypting the message}}- Alice then decrypts the message by applying $D = (AABB)^{4}AA$,   as $DBB=I$. Alice then performs the measurement $M =\sum_{i} |i\rangle\langle i| \otimes I_{c}$ and obtains message  $m'$. The original  message $m$ can then be recovered via $m=m'-l \mod 3$. This protocol can be applied using the $4-$ cycle graph as well, following exactly same procedure as outlined above for $3-$ cycle graph.
\section{ Conclusion} The criteria for generating ordered or periodic QW's by combining  two chaotic QW's has been established. It is shown that if a chaotic QW is obtained by repeatedly applying coins ${A}$ or ${B}$ then one cannot obtain periodic QW by repeatedly applying ${AB}$, ${ABBB}$ or  ${ABBBB}$. Parrando's paradox was seen for the deterministic combination ${AABB}$, i.e., periodic QW was obtained by repeatedly applying ${AABB}$ even when coins ${A}$ and ${B}$ generate chaotic QW's. Our work shows that it is possible to design a quantum periodic signal from two quantum chaotic signals. This also means that its reverse process is also possible, i.e., breaking a periodic quantum signal into two or more quantum chaotic signals. The periodic probability distribution generated from two chaotic ones seen in discrete time QW's on cyclic graphs are of great interest in designing new quantum algorithms, quantum cryptology as well as in development of quantum chaos control theory\cite{whaley}. Finally, there have been recent reports on development of quantum image encryption techniques via chaotic QW's on cyclic graphs\cite{yang}. Our work could provide intriguing possibilities in designing better image encryption protocols. 
\acknowledgments This work was supported by the grants (i) "Josephson junctions with strained Dirac materials and their application in quantum information processing" from Science \& Engineering Research Board (SERB), New Delhi, Government of India, under Grant No. CRG/20l9/006258, and (ii) "Nash equilibrium vs Pareto optimality in N-Player games" SERB MATRICS under Grant No. MTR/2018/000070.
\section{Appendix: Calculation of Lyapunov exponent}  \label{app}
 In   sections~II-V we have used the same definition for chaotic quantum walk as was also used in Ref.~\cite{treganna}, which is if the walker returns to its initial position at $x=0$ after finite number of steps, with probability $1$ then it's periodic else if this probability of return to initial position is never $1$ then it's chaotic. However, another definition of whether the walk is chaotic or not can be determined via the Lyapunov exponent\cite{lyapunov}. A positive value of the Lyapunov exponent, i.e., $\lambda > 0 $ implies the walk is chaotic while if $\lambda=0$ then it's periodic or non-chaotic. We herein below give in short the recipe, see also Ref.~\cite{lyapunov}, for calculating the Lyapunov exponent for cyclic quantum walks and then determine it for our case of a $3-cycle$ quantum walk. In Ref.~\cite{lyapunov}, the chaotic walk is generated via small change in the initial position of the walker. In our study, on the other hand, chaotic quantum walks are generated via unitary operators. To calculate Lyapunov coefficient we start with initial normalized state at  time $t=t_0$ on cyclic graph with $3$ sites, $\left|\Psi\left(t_{0}\right)\right\rangle$.  Then we let the initial state evolve with \[\left|\Psi\left(t\right)\right\rangle=U^{t-t_{0}}\left|\Psi\left(t_{0}\right)\right\rangle,\] with $t > t_{0}.$
Now two cases arise, one wherein the unitary operators,  generate a chaotic walk denoted by $U_c$ and the other wherein they generate a periodic walk, say $U_p$, with period $t_p$.  

Thus, we have for $U=U_p$,  \[\left|\Psi\left(t_{p}\right)\right\rangle=U_{p}^{t_{p}-t_{0}}\left|\Psi\left(t_{0}\right)\right\rangle=\left|\Psi\left(t_{0}\right)\right\rangle,\] i.e., $U_{p}^{t_{p}-t_{0}}=I$, where $I$ is identity matrix. On the other hand, for $U=U_c$,  \[\left|\Psi\left(t\right)\right\rangle=U_{c}^{t-t_{0}}\left|\Psi\left(t_{0}\right)\right\rangle\neq\left|\Psi\left(t_{0}\right)\right\rangle, \forall t. \]  

Thus, one can define a ``distance'' state, i.e., $\left|\Psi_{d}\right\rangle=\left|\Psi\left(t\right)\right\rangle-\left|\Psi\left(t_{0}\right)\right\rangle$ with, $|\Psi_{d}\rangle=0$ for $U=U_{p}$ at $t=t_{p}$ while $|\Psi_{d}\rangle\neq 0$ for $U=U_{c}, \forall t.$ 

This probability distance function $d(t)$ can similarly, to Ref.~\cite{lyapunov}, help us calculate the Lyapunov exponent.  The probability distance function can be expanded as
\begin{eqnarray}
d(t)&=&|\langle\Psi_{d}\mid\Psi_{d}\rangle|=(\langle\Psi(t)|-\langle\Psi(t_{0})|) (\left|\Psi\left(t\right)\right\rangle-\left|\Psi\left(t_{0}\right)\right\rangle),\nonumber\\
&=& |2-2\langle\Psi(t-t_{0})\mid\Psi(t_{0})\rangle|=f(\lambda,(t-t_{0})).
\end{eqnarray}
In the above equation, $\lambda$ is the Lyapunov exponent. The distance function $f(\lambda,(t-t_{0}))$ is bounded by maximum value $2$ and minimum $0$. In chaotic case, $\lambda > 0$ and maximum possible value is $2$, while for periodic case at $t=t_{p}$, $\lambda=0$ and $d=0$, this implies $f(\lambda,(t-t_{0}))=2(1-2^{-\lambda(t-t_{0})})$ and for the Lyapunov exponent one obtains
\begin{eqnarray}
\lambda&=&-\frac{1}{t-t_{0}}\log_{2}|\langle\Psi(t)\mid\Psi(t_{0})\rangle|,\nonumber\\
&=&-\frac{1}{t-t_{0}}(\log_{2}|\langle\Psi(t)\mid\Psi({0})\rangle|+\log_{2}|\langle\Psi(t_{0})\mid\Psi({0})\rangle|,\nonumber\\
\end{eqnarray}
wherein $|\Psi(0)\rangle$ is initial state of walker at time step $t=0$. As the quantum walk is performed in circular path, taking large value of '$t$' is not suitable. Following this we get Lyapunov exponent of process $AAAA...$ as $0.012$, for process $BBBB... $ as $0.085$ and for process $AABB... $ as $0$ for $t-t_{0}=20$ time steps. We also verified that the Lyapunov exponent remains positive for both $AAAA...$ as well as $BBBB... $  for any arbitrary value of $t-t_{0}$ while $\lambda=0$ for  $AABB... $ at time steps $t-t_{0}=20, 40, 60, ...$. Thus, we indeed see that while Lyapunov exponent for  chaotic quantum walks in the $3-cycle$ graph are positive values, for the periodic quantum walk,  Lyapunov exponent vanishes. Similar results can be obtained for $4-$cycle graph wherein also we see for the periodic walk, Lyapunov exponent vanishes while for chaotic walk and finite positive value for the Lyapunov exponent.  

\end{document}